\documentclass[aps,twocolumn,10pt,amsmath,amssymb,nofootinbib,showkeys,showpacs,pra]{revtex4-1}
\usepackage{amsfonts}
\usepackage{amssymb}
\usepackage{graphicx}
\usepackage{subfigure}
\usepackage{color}
\usepackage{ulem}
\usepackage{bbold}
\usepackage{placeins}
\usepackage{soul,xcolor}
\usepackage{epstopdf}
\usepackage{upgreek}
\usepackage{braket}

\setstcolor{red}

\begin{document}

	\title{Geometric phase and neutrino mass hierarchy problem}

	\author{Khushboo Dixit}
	\email{dixit.1@iitj.ac.in}
	\affiliation{Indian Institute of Technology Jodhpur, Jodhpur 342037, India}
	
	\author{Ashutosh Kumar Alok}
	\email{akalok@iitj.ac.in}
	\affiliation{Indian Institute of Technology Jodhpur, Jodhpur 342037, India}
	
	\author{Subhashish Banerjee}
	\email{subhashish@iitj.ac.in}
	\affiliation{Indian Institute of Technology Jodhpur, Jodhpur 342037, India}

	\author{Dinesh Kumar}
	\email{dinesh@uniraj.ac.in}
	\affiliation{University of Rajasthan, Jaipur 302004, India}

	\date{\today} 
	
	\begin{abstract}
We study the geometric phase for neutrinos at various man-made facilities, such as the reactor and accelerator neutrino experiments. The analysis is done for the three flavor neutrino scenario, in the presence of matter and for general, noncyclic paths. The geometric phase is seen to be sensitive to the $CP$ violating phase in the leptonic sector and the sign ambiguity in $\Delta_{31}$. We find that for neutrino experimental facilities where the geometric phase can complete one cycle, all the phase curves corresponding  to different values of $CP$ violating phase,  converge to a single point, called the {\textit{cluster point}}. There are two distinct cluster points for positive and negative signs of $\Delta_{31}$. Thus the geometric phase can contribute to our understanding of the neutrino mass hierarchy problem.
\end{abstract}

%
\vspace{2pc}
%
%
\maketitle
%
%

\section{Introduction}
The phenomenon of neutrino oscillations, which is experimentally well established \cite{Bahcall:2004ut,Eguchi:2002dm,Araki:2004mb,Ashie:2004mr,Michael:2006rx,Abe:2013hdq,Abe:2013fuq}, has emerged as one of the most prominent areas of research over the past decade. Not only it has provided evidence of physics beyond the Standard Model of particle physics but has also provided deep insight into the understanding of mass. The present and future planned facilities in neutrino sector aims at resolving yet many unsolved puzzles, such as observation of $CP$ violation in the leptonic sector, resolving mass hierarchy problem, probing sterile neutrinos, resolving Dirac vs Majorana neutrinos, probing ultra high energy neutrinos, dark matter searches. Recently neutrinos have been suggested as a tool to probe  foundational issues in quantum mechanics \cite{Blasone:2009xk,Blasone:2014jea,Alok:2014gya,Banerjee:2015mha,Naikoo:2017fos}.

The $CP$ violation in the quark sector has been measured and well established in the standard model of particle physics \cite{pdg}. The $CP$ violation in the quark mixing matrix can arise due to the complex Yukawa couplings. A relative $CP$ violating phase in the vacuum expectation values  of the Higgs fields can also give rise to the $CP$ violation in the quark sector. An analogous mechanism is expected to induce $CP$ violation in the leptonic sector which has yet not been observed. The neutrino oscillation experiments provide a platform to measure the $CP$ violation in the leptonic mixing matrix known as the PMNS matrix  \cite{Dick}. 	The presently running and planned experiments in neutrino sector also aim to resolve the mass hierarchy problem, i.e., whether the mass appearing in $\nu_3$-eigenstate of neutrino is the lightest or the heaviest of the three masses.
In this work we study these issues in the context of geometric phase (GP) in neutrino oscillations.

The concept of GP exists in the standard Pontecorvo formulation of  neutrino oscillations \cite{Blasone}. The adiabaticity condition needs to be satisfied in order to have Berry phase. This condition is provided by varying the matter density. The Berry phase in the context of two flavor neutrino oscillations was calculated in Ref.~\cite{Blasone}. The possibility of observation of Berry phase in long baseline neutrino oscillation experiments was discussed in \cite{He}. In order to extract Berry phase, one needs to vary the oscillation distance $L$ while maintaining the cyclicity condition \cite{He}. Thus one has to measure this phase after the state has undergone a complete cycle with time period $T$. In neutrino oscillations, this period is relatively long so, one needs to place the detector at sufficiently large distance from the source to observe the Berry phase. The generalization of GP to noncyclic paths was made shortly after Berry's discovery involving closed cyclic paths, see for example \cite{wilczek}. The noncyclic phase could be easier to observe because the cyclicity constraint is removed \cite{Xiang}, where the analysis was done for vacuum case only. Here we take this up further by including earth's matter effect and inputs from ongoing neutrino oscillation experimental setups. Recently, attempts have been made towards establishing the Dirac or Majorana nature of neutrinos using noncyclic GP  \cite{Dajka,Capolupo}.
 In \cite{pmehta} the topological phase in two-flavor neutrino oscillations was discussed by using the  analogy between the two-flavor state and the polarization states in optics. In \cite{Johns:2016wjd} neutrino geometric phases in high-density environments were studied.

GP has a rich and interesting history.  It was introduced by Berry \cite{berry}, in the quantum scenario, for the case of cyclic adiabatic evolution. The connection of this work to the earlier work of Pancharatnam \cite{pancham} related to the interference of polarized light was made in \cite{nityananda}. Berry's phase was shown to be a consequence of the holonomy in a line bundle over parameter space \cite{simon}, bringing out the geometric nature of the phase. Generalization of Berry's work to non-adiabatic evolution was made in \cite{aharanov} and to the case of non-cylic evolution in \cite{samuel}. 
A kinematic approach to GP for mixed states undergoing nonunitary evolution was proposed in \cite{tong}. In the context of open quantum systems, GP has been studied for a variety of systems, see for example, \cite{zenardi,villar,sb1,sb2,sb3}. In \cite{lidar}  Abelian and non-Abelian geometric phases for open quantum systems was considered. GP has also been studied in the context of neutral meson systems to probe $CPT$ violation  \cite{Capolupo1}.

In this work we study non-cyclic GP in the context of three flavor neutrino oscillations in the presence of matter and $CP$ violating effects. We discuss how GP can be used to distinguish between the normal and inverted hierarchy of neutrino masses. We analyze GP in the context of two  reactor neutrino experiments, Daya-Bay \& RENO and two accelerator experiments, T2K \& NO$\nu$A. We find that the GP in  T2K, Daya-Bay and RENO experimental set-ups can distinguish effects of normal and inverted mass hierarchy for all values of $CP$ violating phase.  We also discuss the possibility of measurement of geometric phase at the currently running neutrino physics experimental facilities.

The paper is arranged as follows. After the introduction, we provide theoretical expressions for geometric phase, both in vacuum and in presence of matter. In Sec. III,  we present our results for different experimental set-ups.  In Sec. IV, we show that the geometric phase can be expressed in terms of neutrino survival and oscillation probabilities. Finally, in Sec. V, we present our conclusions.

\section{Noncyclic geometric phase}
Let us consider a state $\ket{\chi(0)}$  which evolves to a state $\ket{\chi(t)}$,
after a certain time $t$. Then the scalar product 
\begin{equation}
\bra{\chi(0)} \exp\left[\frac{i}{\hbar} \int_{0}^{t} \mathrm{\left<E\right>}(t')dt'\right] \ket{\chi(t)},
\end{equation}
can be written as $r\,e^{i\beta}$, where $r$ is real number and angle $\beta$ is the noncyclic phase due to the evolution from $\ket{\chi(0)}$ to $\ket{\chi(t)}$ \cite{Xiang}. The experimentally measurable quantities in neutrino oscillation experiments  are survival and oscillation probabilities i.e., $P_{\nu_{i} \rightarrow \nu_{j}} = |\left<\nu_{j}|\nu_{i}\right>|^2$. The phase factor in the quantity $\left<\nu_{j}|\nu_{i}\right>$ has two parts, dynamical ($\xi$) and geometric ($\beta$),
\begin{equation}
\left<\nu_{j}|\nu_{i}\right> = r e^{-i \beta} e^{-i \xi}.
\end{equation}
Since the dynamical phase depends on the energy of the system, it can be removed using gauge transformation, such as $\ket{\psi} \rightarrow \ket{\tilde{\psi}} = e^{i \xi} \ket{\psi}$ where $\xi = \frac{1}{\hbar}\int_{0}^{t} E(t') dt'$. However, the geometric part of the total phase still exists in the amplitude $\left<\nu_{j}|\nu_{i}\right>$ even after the gauge transformation \cite{Xiang,samuel}. As both the geometric phase and probabilities are functions of mixing angle $\theta$ and mass square difference $\Delta$, the geometric phase can be expressed in terms of survival and oscillation probabilities and their average values. This is shown explicitly for the two flavor case in Sec. IV below.\par  
We now consider three flavor oscillating neutrino states, 
$ \ket{\nu_\alpha(0)} = U_{\alpha a}\ket{\nu_a}$, 
where $\ket{\nu_\alpha(0)}$ is initial flavor state ($\alpha$ = $e$, $\mu$, $\tau$) and $\ket{\nu_a}$ are mass eigenstates ($a = 1,\,2,\,3$). $U_{\alpha a}$ are the elements of the PMNS mixing matrix $U$ given by 
{\small
	\begin{equation}
	U = \begin{pmatrix}
	c_{12}c_{13}                   &               s_{12}c_{13}                    &    e^{i\delta}s_{13} \\
	-s_{12}c_{23}-c_{12}s_{23}s_{13}e^{-i\delta} &  c_{12}c_{23}-s_{12}s_{23}s_{13}e^{-i\delta}  &     s_{23}c_{13} \\
	s_{12}s_{23}-c_{12}c_{23}s_{13}e^{-i\delta}  &  -c_{12}s_{23}-s_{12}c_{23}s_{13}e^{-i\delta} &     c_{23}c_{13}
	\end{pmatrix}.
	\end{equation}
}

Here $c_{ij}=\cos \theta_{ij}$, $s_{ij}=\sin \theta_{ij}$ and $\delta$ is the $CP$ violating phase.

The time evolution of $\ket{\nu_\alpha}$ in the flavor basis is given by
\begin{equation}
\ket{\nu_\alpha(t)} =U\ U_m(t)\ U^{-1} \ket{\nu_\alpha(0)} 
=U_f(t) \ket{\nu_\alpha(0)}.
\end{equation}
\begin{equation}
U_m = \begin{pmatrix}
e^{-i\omega_1t} &        0 &                  0 \\
0      &  e^{-i\omega_2t} &          0 \\
0      &        0 &           e^{-i\omega_3t}
\end{pmatrix}
\end{equation}  

and $U_f(t)$ = $U\ U_m(t)\ U^{-1}$.

To calculate the noncyclic phase generated during the time evolution of, say $\ket{\nu_\mu(0)}$ to  $\ket{\nu_\mu(t)}$, we define a new state $\ket{\tilde{\nu}_\mu(t)}$, given by 
\begin{equation}
\ket{\tilde{\nu}_\mu(t)} = \exp\left[ i \int_{0}^{t} \mathrm{\left<E\right>} (t')dt'\right] \ket{\nu_\mu(t)},
\end{equation} 
where 
$\mathrm{\left<E\right>} (t') = \bra{\nu_{\mu}(t')} i \partial_{t} \ket{\nu_{\mu}(t')}$,    
and
\begin{equation}
\ket{\nu_{\mu}(t')}= U_{f12}(t')\ket{\nu_{e}} +  U_{f22}(t')\ket{\nu_{\mu}} +  U_{f32}(t')\ket{\nu_{\tau}},
\end{equation}
with $\langle \nu_{\mu}(0)| \nu_{\mu}(t')\rangle = U_{f22}(t') $,
$U_{fab}(t')$ are a$\times$b elements of the evolution operator $U_{f}$, ($a, b = 1, 2, 3$).
Then the scalar product is given by
\begin{equation}
\begin{aligned}
\langle\nu_{\mu}(0) | \tilde\nu_{\mu}(t)\rangle = \exp\left[i \int_{0}^{t} \mathrm{\left<E\right>} (t')dt'\right]~U_{f22}(t')
\end{aligned}
\begin{aligned}
\equiv r_{\mu\mu} e^{i \beta_{\mu\mu}},
\end{aligned}
\end{equation}
so that the noncyclic phase due to evolution from $\ket{\nu_\mu(0)}$ to $\ket{\nu_\mu(t)}$ will be the
phase angle $\beta_{\mu\mu}$.  Similarly, it is possible to compute the other inner products, such as, $\langle\nu_{\mu}(0) | \tilde\nu_{e}(t)\rangle$ and their corresponding noncyclic phases.

\subsection{Non cyclic geometric phase in vacuum} 
Using the definition of non-cyclic GP from the previous section, the theoretical expressions of GP, such as $\beta_{ee}$, $\beta_{e \mu}$, $\beta_{\mu \mu}$ and $\beta_{\mu e}$, are 	\cite{Xiang}
\begin{widetext}
	\begin{eqnarray}
	\beta_{ee}&=& (2s_{12}^2 c_{13}^2 + 2q s_{13}^2 -1) (\phi \frac{L}{c})\nonumber\\
	&&+\tan^{-1} \frac{\cos(2\theta_{12}) c_{13}^2 \sin(\phi\frac{L}{c})-s_{13}^2 \sin[(2q-1)\phi\frac{L}{c}]}{c_{13}^2 \cos\phi\frac{L}{c}+s_{13}^2 \cos[(2q-1)\phi\frac{L}{c}] )},\label{gp1}
	\\
	\beta_{e \mu} &=& \big[\cos (2\theta_{12})(c_{23}^2 - s_{13}^2 s_{23}^2) + c_{13}^2 s_{23}^2 (2q -1) - s_{13}
	\sin (2 \theta_{12}) \sin (2 \theta_{23}) \cos (\delta)\big] (\phi \frac{L}{c})\nonumber\\
	&&+\tan^{-1}\frac{s_{13}s_{23}[s_{12}^2 \sin (\phi\frac{L}{c}+\delta)-c_{12}^2 
		\sin (\phi\frac{L}{c}-\delta) -\sin \psi_{+}]-c_{23} \sin (\phi\frac{L}{c})\sin (2\theta_{12})}
	{\sin (\theta_{13})\sin (\theta_{23})[cos \psi_{+} 
		-s_{12}^2 \cos (\phi\frac{L}{c}+\delta)-c_{12}^2 \cos (\phi\frac{L}{c}-\delta)]}, \label{gp2}
	\\
	\beta_{\mu\mu} &=& \big[\cos (2\theta_{12})(c_{23}^2- s_{13}^2 s_{23}^2)+c_{13}^2 s_{23}^2 (2q-1)-s_{13} \sin (2\theta_{12})
	\sin (2\theta_{23}) \cos \delta\big](\phi \frac{L}{c})
	\nonumber\\
	&&+\tan^{-1}\frac{s_{13} \sin (2\theta_{12}) \sin (2\theta_{23}) \cos \delta \sin (\phi\frac{L}{c})-(c_{23}^2-s_{23}^2 
		s_{13}^2) \cos (2\theta_{12})\sin (\phi\frac{L}{c})-c_{13}^2 s_{23}^2 \sin \Gamma
	}{(c_{23}^2+s_{13}^2 s_{23}^2) \cos (\phi\frac{L}{c})+c_{13}^2 s_{23}^2 \cos \Gamma},~~~~~~ \label{gp3}
	\\
	\beta_{\mu e}&=&(2 s_{12}^2 c_{13}^2 +2qs_{13}^2-1)(\phi\frac{L}{c})\nonumber\\
&&	+\tan^{-1}\frac{s_{13}s_{23}[s_{12}^2 \sin (\phi\frac{L}{c}-\delta)-c_{12}^2 \sin (\phi\frac{L}{c}+\delta) 
		-\sin \psi_-]-c_{23} \sin (\phi\frac{L}{c})\sin (2\theta_{12})}{\sin (\theta_{13})\sin (\theta_{23})[\cos \psi_- 
		-s_{12}^2 \cos (\phi\frac{L}{c}-\delta)-c_{12}^2 \cos (\phi\frac{L}{c}+\delta)]},  \label{gp4}       
	\end{eqnarray}
\end{widetext}
where
$ q =(\omega_3-\omega_1)/(\omega_2-\omega_1) = (\Delta_{32}/\Delta_{21})+1$;
$ \phi = \Delta_{21}/4 E \hbar$;
$ \Gamma = (2q-1)(\phi L/c)$ and
$ \psi_{\pm} = (2q-1)(\phi L/c)\pm\delta$.
Further, $L$ is the propagation length and $E$ is the neutrino energy. For the cyclic case scenario, $q$ would be a rational number with the cyclic period $t = 2\pi/(\omega_1 -\omega_2)$ \cite{Blasone,Xiang}. 

\begin{figure}[h!] 
\centering
\begin{tabular}{c}
\includegraphics[width=62mm]{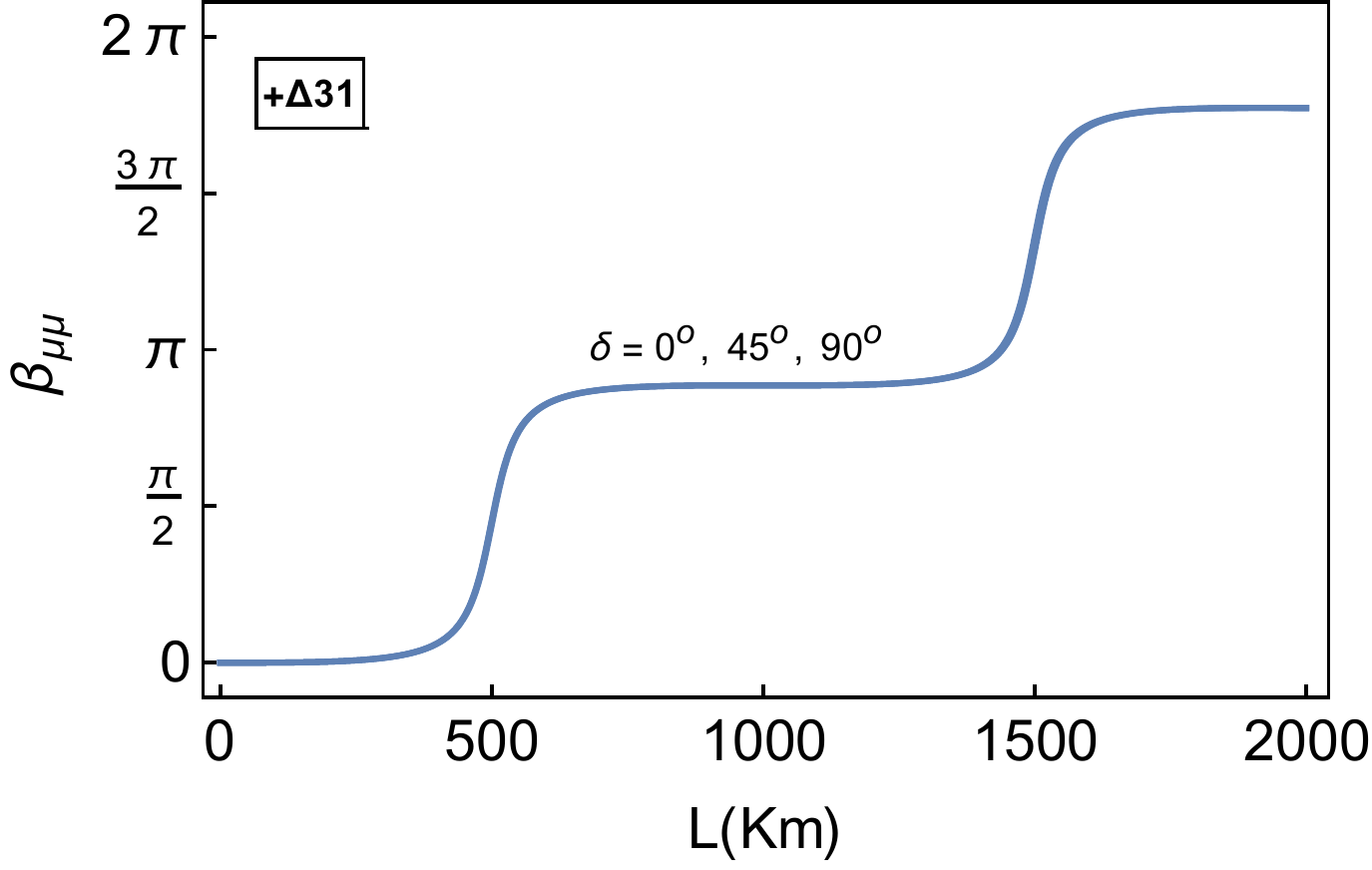}\\
\includegraphics[width=62mm]{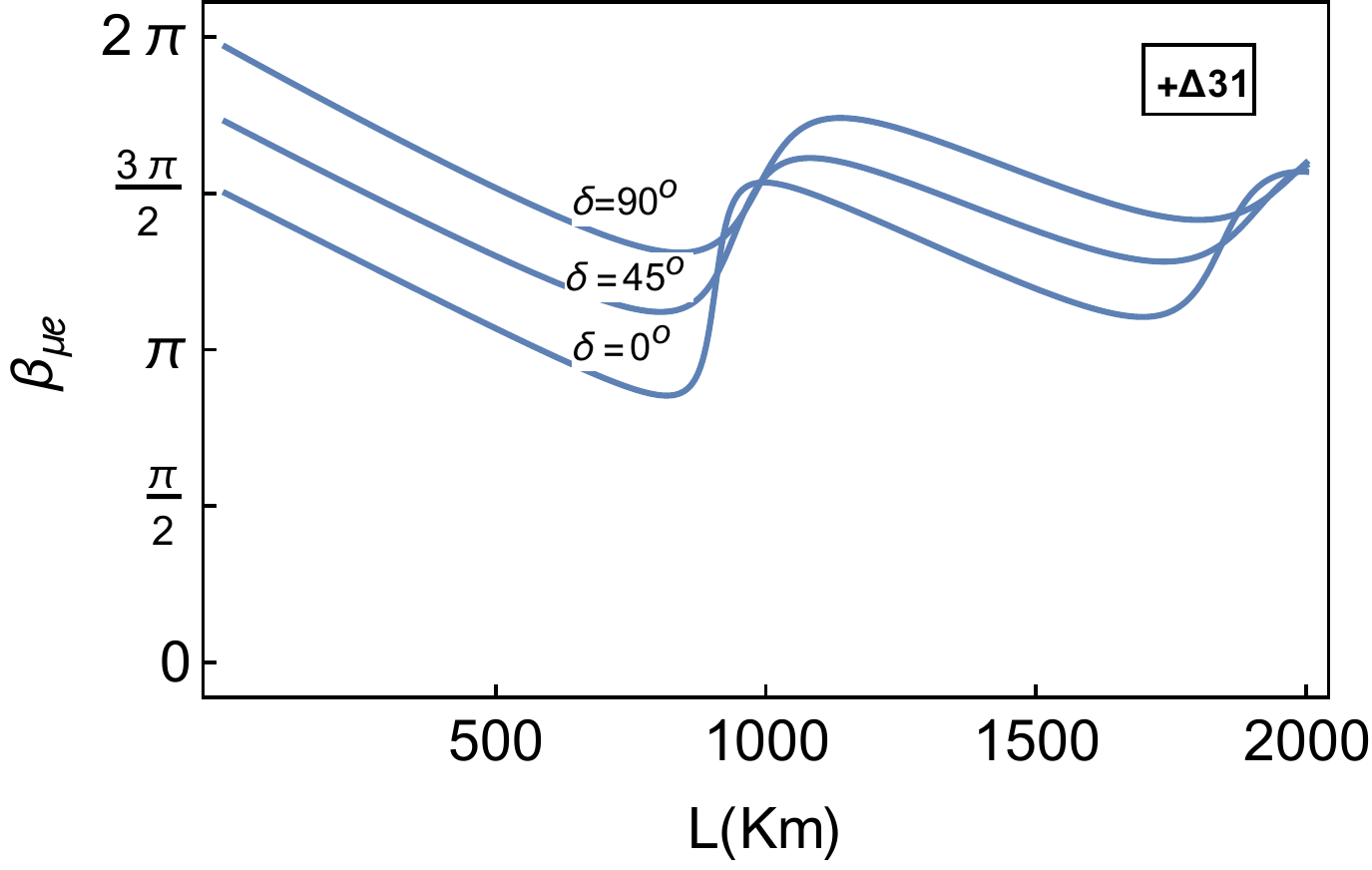}\\
\end{tabular}
\caption{The top and bottom figures show the variation of $\beta_{\mu \mu}$ and $\beta_{ \mu e}$ with $L$ (in Km), respectively in vacuum for different values of $\delta$. Here neutrino energy $E$ = 1 GeV.}
\label{fig:main}
\end{figure}

In our analysis, the values of mixing angles, $\theta_{ij}$, and mass square 
differences, $\Delta_{ij} = m_i^2 - m_j^2$ ($m_i$ and $m_j$ are the masses of the neutrino mass eigenstates $\nu_i$ and $\nu_j$, respectively),  used are $\theta_{12} = 33.48^{o}$, $\theta_{13} = 8.5^{o}$, $\theta_{23} = 42.3^{o}$, 
$\Delta_{21} = 7.5 \times 10^{-5} eV^{2}$, $\Delta_{31}\, (\approx \Delta_{32}) = \pm 2.457 \times 10^{-3} eV^{2}$. The plots of $\beta_{\mu \mu}$ and $\beta_{\mu e}$ with respect to $ L$ 
(for two cycles) for neutrino energy $E$ = 1 GeV for different values of $\delta$ are shown in Fig.~\ref{fig:main}. As can be seen from the figure, for $\beta_{\mu \mu}$, there is mild dependence on $\delta$ whereas $\beta_{ \mu e}$ is sensitive to $\delta$. Similarly, we find that $\beta_{ e \mu}$ is sensitive to the $CP$ phase whereas $\beta_{e e}$ is not.  Therefore in order to study the effect of $CP$ violating phase on GP, we consider $\beta_{e \mu}$ and $\beta_{\mu e}$ in this work.

\subsection{Non cyclic geometric phase in the presence of earth matter effects}	 

In order to calculate non-cyclic GP in the presence of matter, we make use of the formalism given in \cite{T01,Tommy,T02}. In this formalism, the evolution operator which evolves an initial flavor state, in mass eigenstate basis and flavor state basis in a constant matter density are given by
\begin{eqnarray}
U_m(L) &=& \phi \sum_{a=1}^{3} \frac{e^{-iL \lambda_a} }{3\lambda_a^2 +{c_1}} [(\lambda_{a}^2+c_1)I + \lambda_a T + T^2],\\
U_f(L) &=& \phi \sum_{a=1}^{3}  \frac{e^{-iL \lambda_a}}{3\lambda_a^2 +{c_1}} [(\lambda_{a}^2+c_1)I + \lambda_a \tilde T + 
\tilde T^2],
\end{eqnarray}
respectively.
The $T$ matrix has the form
\begin{widetext}
	\begin{equation}
	\mathbf{T} =\begin{pmatrix}
	A U_{e1}^{2} - \frac{1}{3} A + \frac{1}{3} ( E_{12} + E_{13} ) & A U_{e1} U_{e2} & A U_{e1} U_{e3} \\ A U_{e1} U_{e2} & A  U_{e2}^{2} - \frac{1}{3} A + \frac{1}{3} ( E_{21} + E_{23} ) & A U_{e2} U_{e3}   \\ A U_{e1} U_{e3}  &  A U_{e2} U_{e3}  & A U_{e3}^{2} - \frac{1}{3} A + \frac{1}{3}(E_{31} + E_{32})
	\end{pmatrix},
	\end{equation}
\end{widetext}
with $\phi = e^{\frac{-i}{3} L Tr{\mathcal{H}_m}}$, $T=\mathcal{H}_m - Tr[\mathcal{H}_m]/3$, $c_1= {\rm det}T ~Tr (T^{-1})$ and $\lambda_{a}$ are eigenvalues of $T$, matter potential  $A = \pm \sqrt{2}\,G_{F}\,n_e$ (+ for neutrinos, - for anti-neutrinos), $G_{F}$ is Fermi constant, $n_e$ is the electron number density, $E_{ij} = E_i - E_j$ and $\tilde{T}= U T U^{-1}$.

If neutrinos travel through a series of constant matter densities, say $A_{1}$, $A_{2}$, $A_{3}$,...$A_n$, and evolution operators corresponding to these matter densities are $U_{1}$, $U_{2}$, $U_{3}$,.....$U_n$ then $U' = U_{1} U_{2} U_{3}....U_n$.
For Earth's mantle-core-mantel step function density profile, with $U_{1}$ and $U_{2}$ corresponding to mantle and core of the Earth respectively, $U'$ will become
$U' = U_{1} U_{2} U_{1}$. The changes in the evolution operator translate into corresponding changes in the noncyclic phases.

\begin{figure}[h!] 
	\centering
	\begin{tabular}{c}
		\includegraphics[width=75mm]{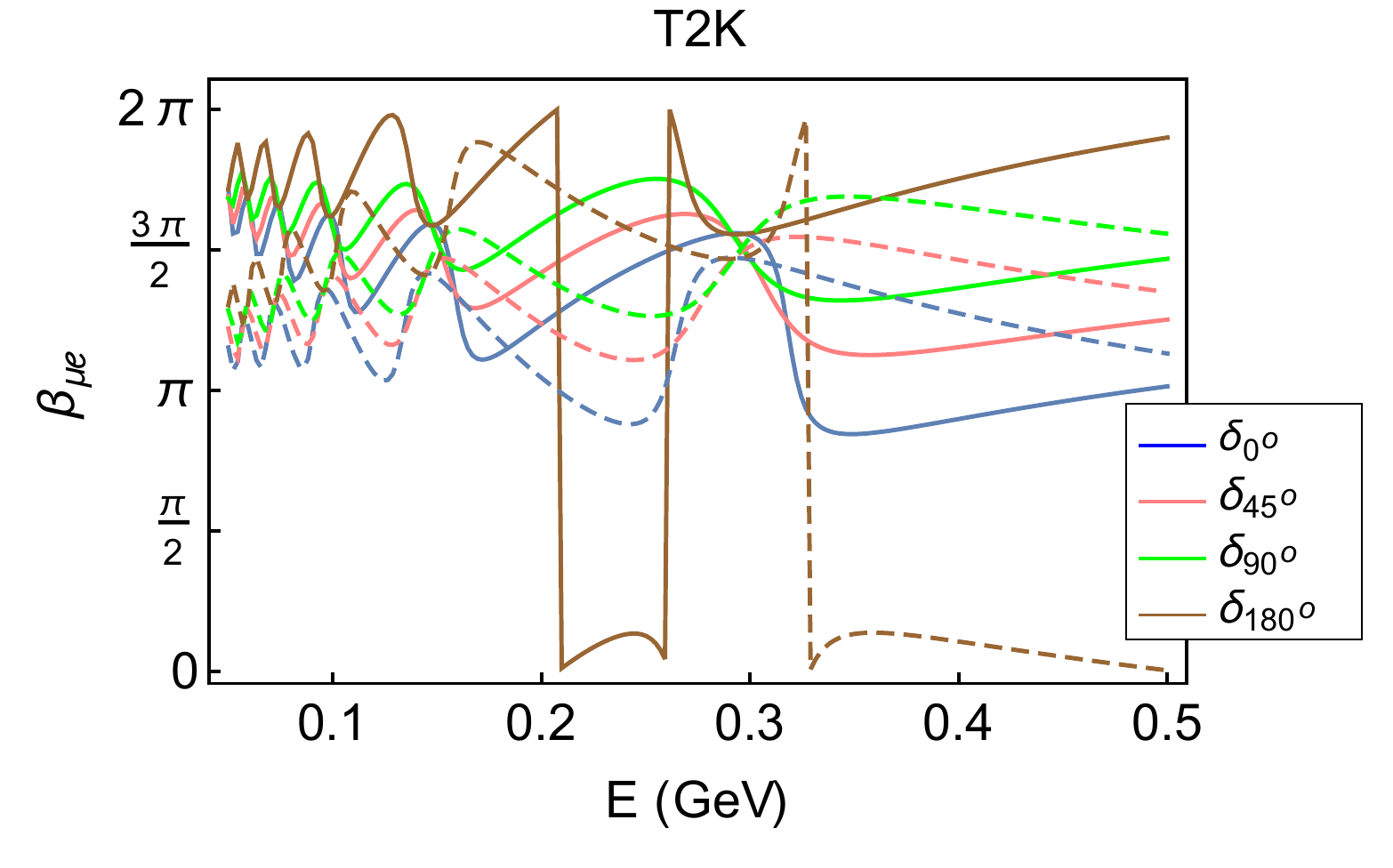}\\
		\includegraphics[width=75mm]{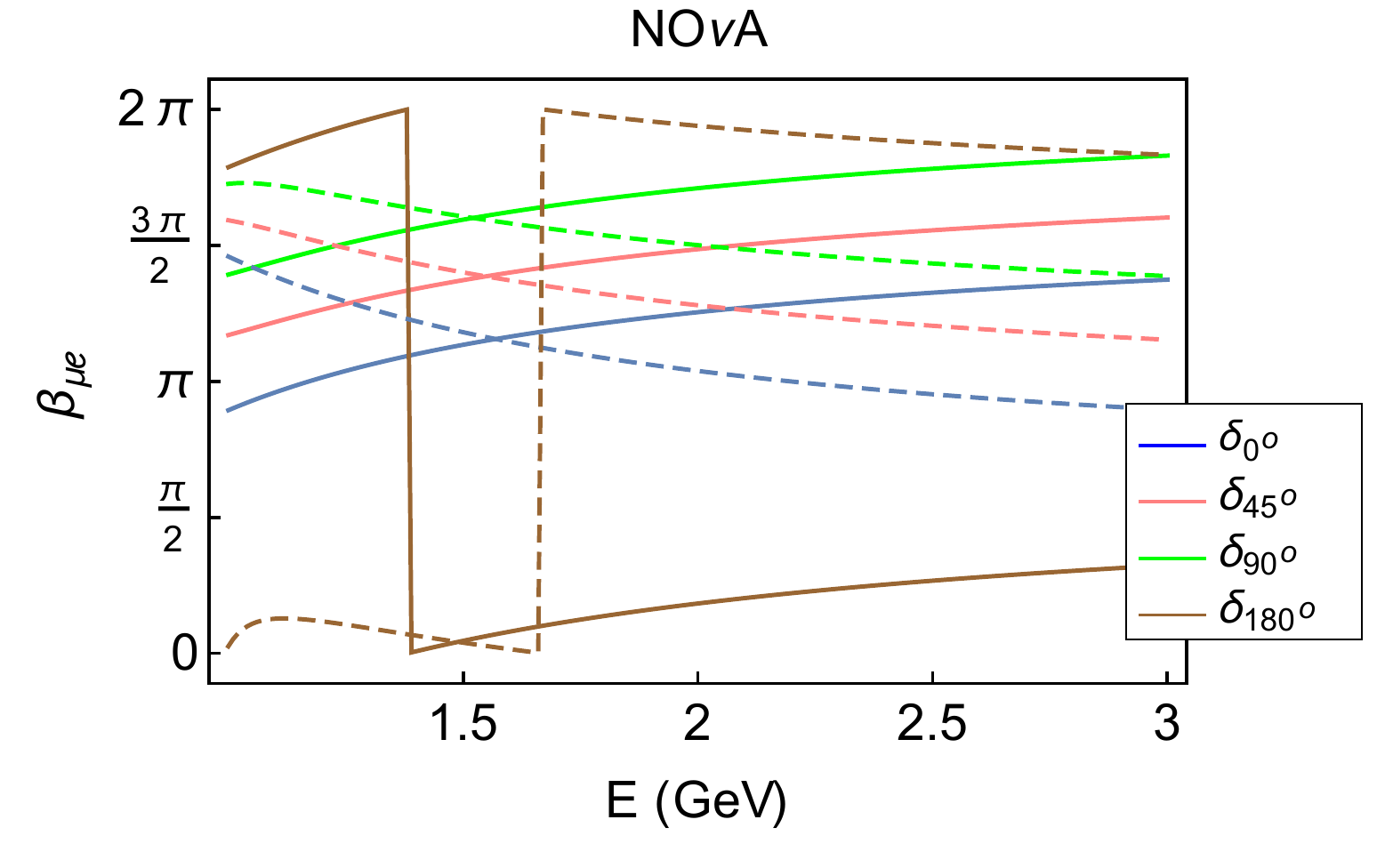}\\
		\includegraphics[width=75mm]{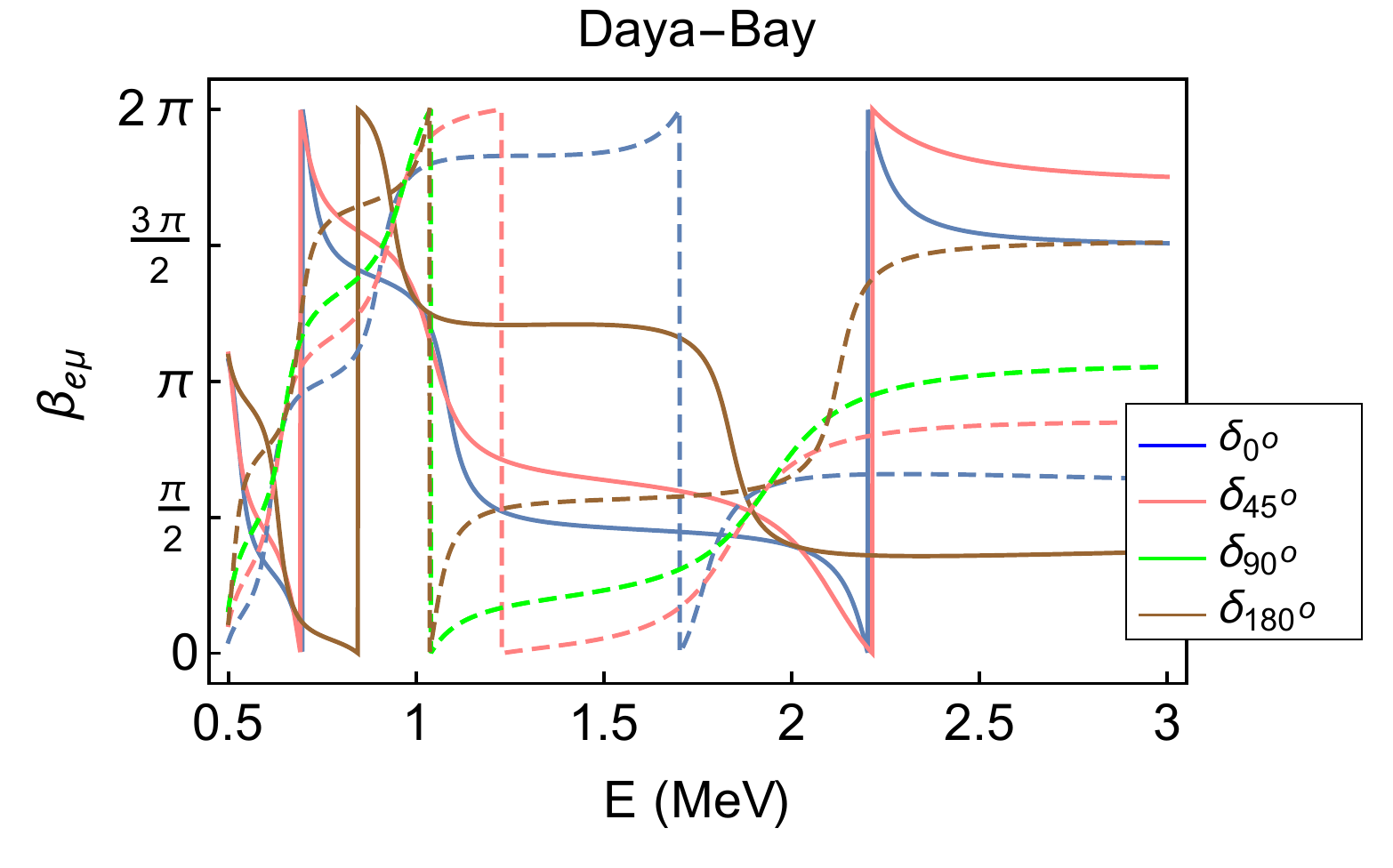}\\
		\includegraphics[width=75mm]{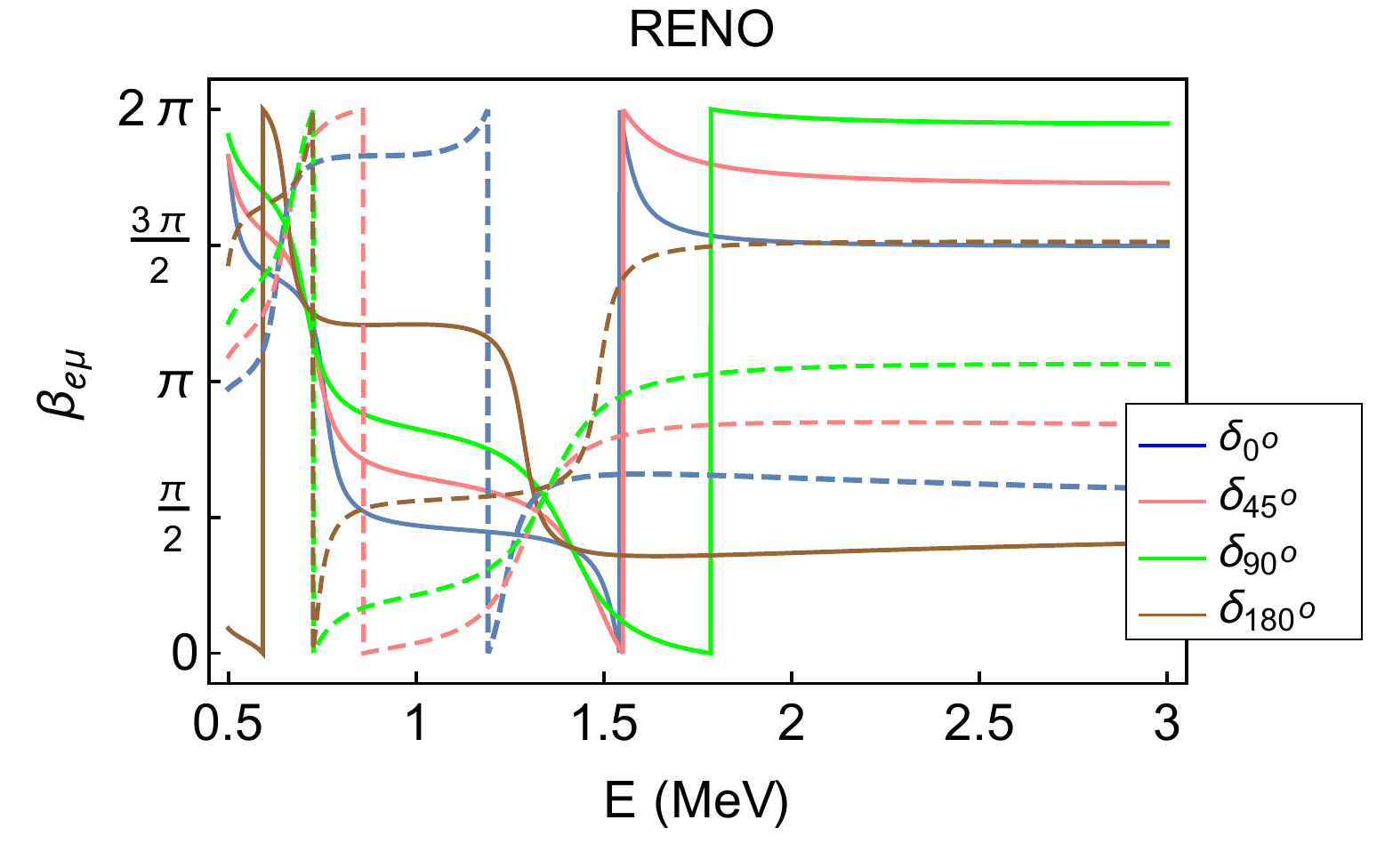}\\
	\end{tabular}
	\caption{First and second figures from top depict $\beta_{\mu e}$ vs $E$ (GeV) plots for T2K and NO$\nu$A experimental set-ups, respectively whereas the third and fourth figures show the variation of $\beta_{e \mu}$ with $E$ (MeV) for Daya-Bay and RENO set-ups, respectively for different $CP$-violating phases $\delta=0$ (blue), $\delta=\frac{\pi}{4}$ (pink), $\delta=\frac{\pi}{2}$ (green) and $\delta=\pi$ (brown). Solid and dashed lines in these plots correspond to positive and negative signs of $\Delta_{31}$, respectively.}
	\label{fig:geophs}
\end{figure}

\section{Results and discussions}
We now study GP  for neutrinos produced at various man made facilities such as the reactor  and 
accelerator neutrino experiments.  For reactor neutrinos, we consider Daya-Bay and RENO experimental 
set-ups whereas for accelerator neutrinos, we present results for NO$\nu$A and T2K.  
Daya-Bay is a China based reactor neutrino experiment\cite{An,Balantekin} which uses a cluster of six nuclear reactors to 
get $\bar{\nu}_e$ anti-neutrinos with energy range of MeV order and three detectors are placed at distance of units of Km. One far detector is placed at approximately  2 Km distant from the source. This baseline has to face the constant rock matter density with matter potential A = $ \rm  -1.01 \times 10^{-13}$ eV (negative sign due to $\bar{\nu}_{e}$).  RENO is also a reactor $\bar{\nu}_{e}$-experiment. Its baseline is 1.4 Km with energy range in units of MeV. Since initial flavor state in these experiments are $\bar{\nu_e}$, therefore we analyze
$\beta_{e \mu}$ component of the GP.

Accelerator neutrino-experiment NO$\nu$A (NuMI Off-Axis $\nu_e$ Appearance) experiment uses NuMi-beam of accelerator $\nu_{\mu}$-neutrinos based at Fermilab \cite{Patterson,Adamson}. The detector is 810 Km (baseline) distant from the source of $\nu_{\mu}$ neutrinos in the energy range of 1-10 GeV. Its aim is to measure the small mixing angle $\theta_{13}$, the neutrino-mass hierarchy and the $CP$-violating phase. This baseline passes through the Earth's crust which has a constant matter density $\rm 1.7\times 10^{-13}$ eV. T2K is an off-axis experiment \cite{Abe:2013hdq,Abe:2013fuq} which uses a $\nu_{\mu}$- neutrino beam from Tokai to Kamioka with energy-range of approximately 100 MeV to 1 GeV. It has 295 Km long baseline and passes through a matter density of $\rm  1.01\times 10^{-13}$ eV. For accelerator neutrinos, we consider $\beta_{ \mu e}$ component of GP.

Fig.~\ref{fig:geophs} shows the variation of $\beta_{e \mu}$ (for Daya-Bay and RENO) and $\beta_{\mu e}$ (for T2K and NO$\nu$A)   with neutrino energy, $E$, for different values of $CP$ violating phase $\delta$ for initial anti electron neutrino and muon neutrino beams, respectively.
It can be seen from the figure that T2K, Daya-Bay and RENO have  neutrino energies and baselines such that the GP can complete at least one cycle. This is because  of the factor 
$\textit{(2q-1)$\phi$}= 
(\frac{\Delta_{32}+\Delta_{31}}{4 E \hbar c})= (\frac{2\Delta_{32}+\Delta_{21}}{4 E \hbar c})$, in Eqs. (\ref{gp2}) and (\ref{gp4}),  which 
plays a leading role in the oscillatory nature of the GP. In accordance with this factor, the value of $L/E$, corresponding to one cycle, should be $\sim$ 0.99 Km/MeV  for $+\Delta_{31}$ and $\sim$ 1.03 Km/MeV for $-\Delta_{31}$. While, the value of $L/E$ is 2950 Km/GeV (2.95 Km/MeV) for T2K (at $L$ = 295 Km, $E$ = 100 MeV), 2 Km/MeV for Daya-Bay (at $L$ = 2 Km, $E$ = 1 MeV)  and 1.4 Km/MeV for RENO 
(at $L$ = 1.4 Km, $E$ = 1 MeV) which is greater than the required $L/E$ value for one cycle,  the value of $L/E$ for NO$\nu$A (at $L$ = 810 Km, $E$ = 1 GeV) $\sim$ 810 Km/GeV
(0.81 Km/MeV) is not enough to have at least one cycle. These $L/E$ values for specific experimental set-ups are calculated at their corresponding lowest energy-values of neutrinos and the baseline length.
\FloatBarrier

Further, we find that for $L/E$ corresponding to a cycle, all GP curves corresponding  to different values of $CP$-phase,  converge to a single point. We call this point as {\textit{cluster point}}. This is so because in the limit $ {\phi \frac{L}{c} \to \frac{2\pi}{(2q-1)}}$, i.e., where one cycle of oscillating feature of GP gets completed, $\beta_{e \mu}$ remains approximately constant for every $CP$ violating  phase $\delta$.
An interesting thing to observe here is that there are two distinct cluster points corresponding to + and - signs of $\Delta_{31}$.  This would thus help in removing the sign ambiguity in $\Delta_{31}$. Therefore if we choose proper energy range, for the fixed baseline length, such that $L/E$ corresponds to a cycle of the GP, then one can disentangle effects of + and - signs of $\Delta_{31}$. For example, for the case of Daya-Bay experiment (L = 2Km), we are getting these cluster points at E = 1 MeV and 2 MeV. So one has to tune the energy value of anti-neutrinos at these values and by measuring the GP-value, the effect of + and - signs can be disambiguated.

\begin{figure}[h!] 
	\centering
	\begin{tabular}{c}
		\includegraphics[width=70mm]{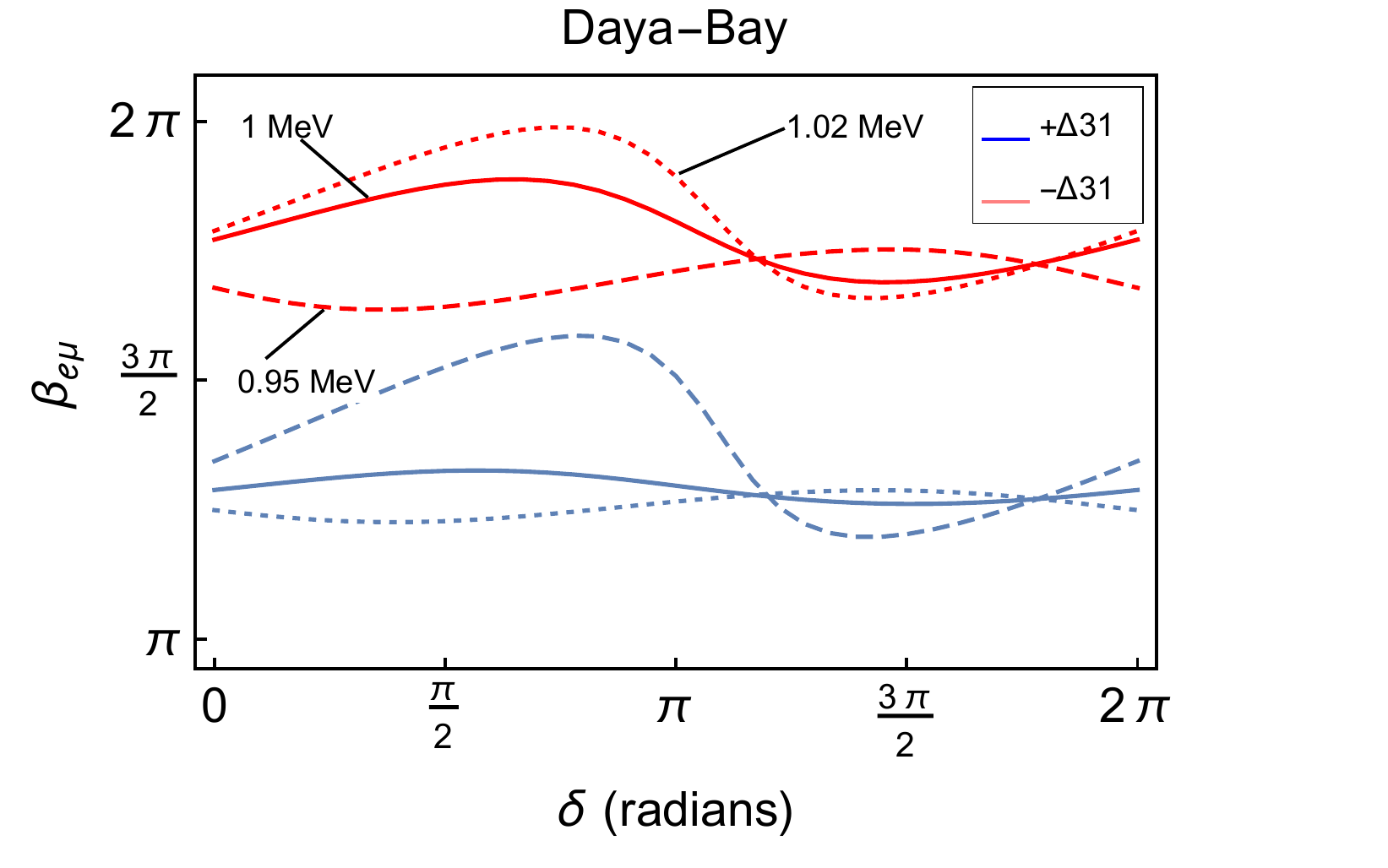}\\
		\includegraphics[width=70mm]{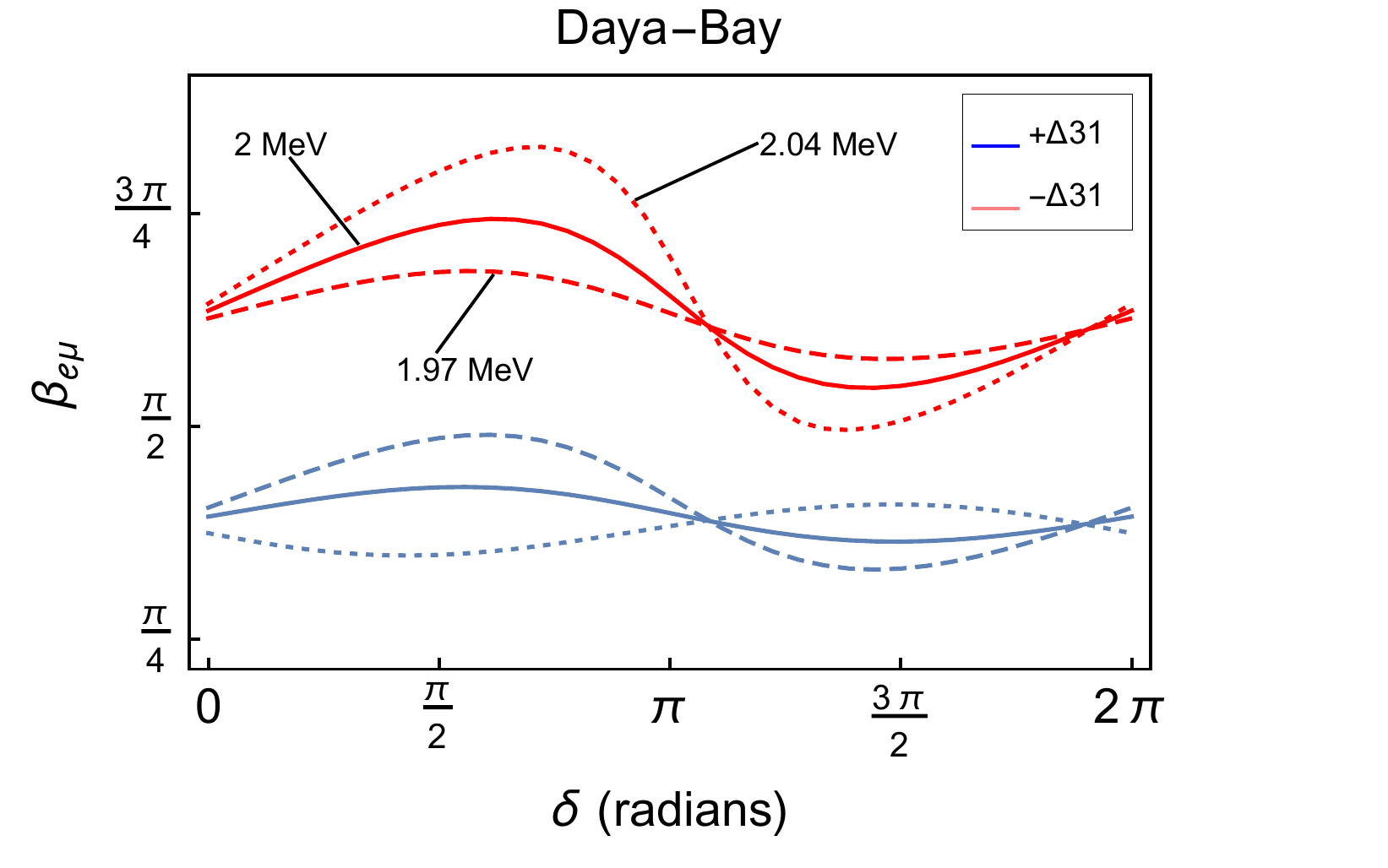}\\
		\includegraphics[width=70mm]{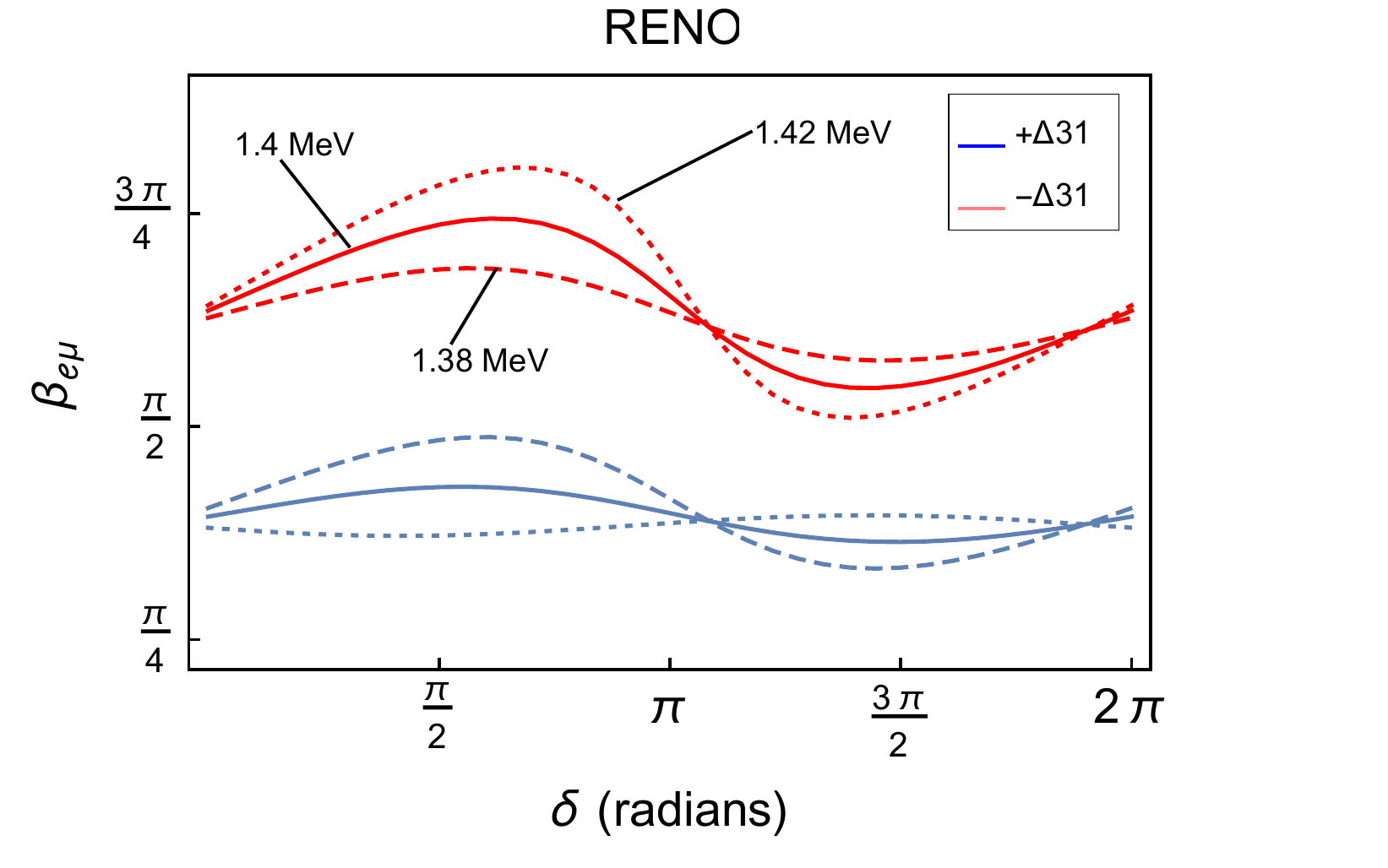}\\
		\includegraphics[width=70mm]{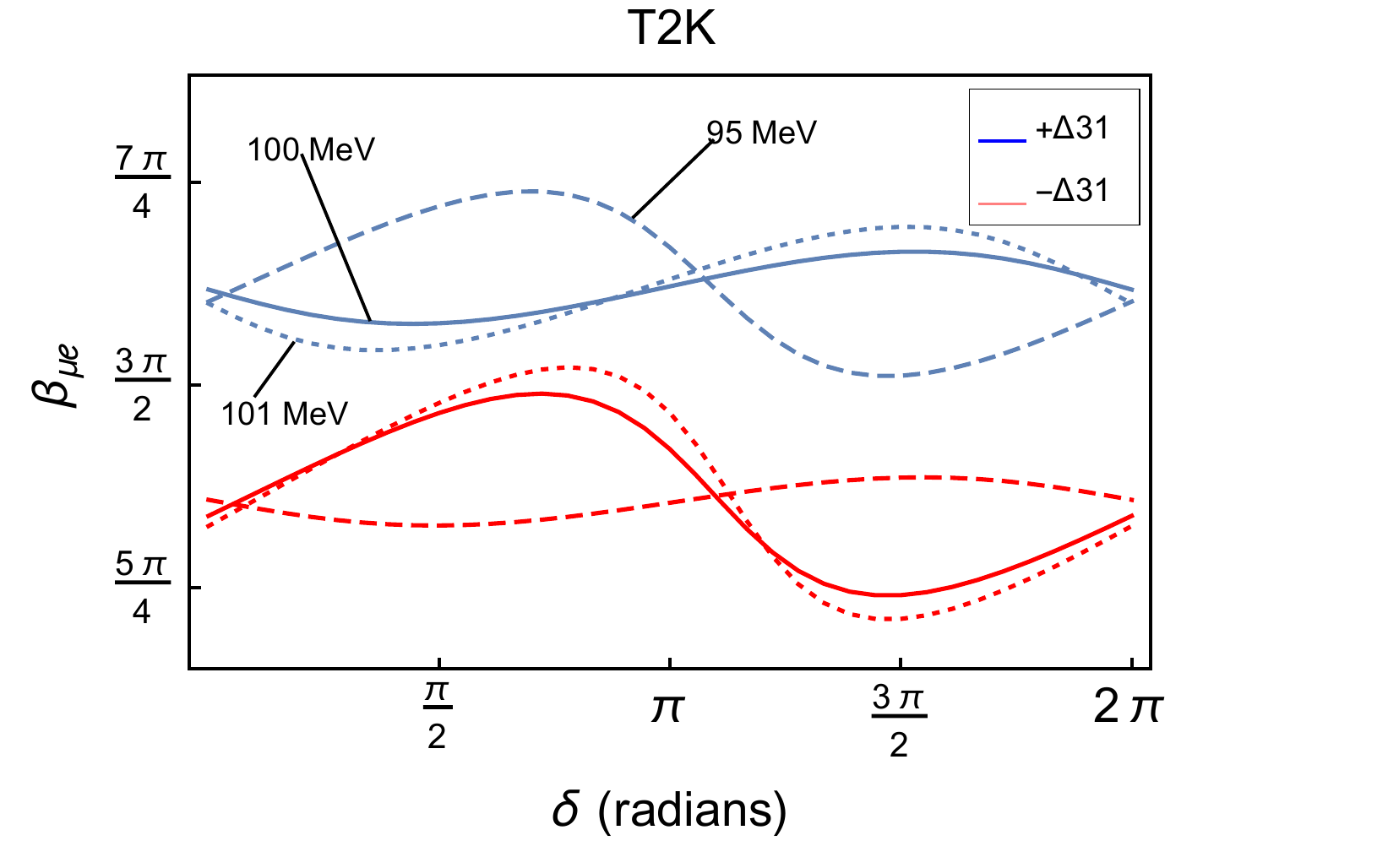}\\
	\end{tabular}
	\caption{From top, the first and second figures show $\beta_{e \mu}$ vs $\delta$ plots for Daya-Bay experimental set-up. The third and fourth figures show variation of $\beta_{e \mu}$ and $\beta_{ \mu e}$ with $\delta, $ for RENO and T2K experimental set-ups, respectively. Solid curves in these plots correspond to that specific value of neutrino energy at which a cycle of GP gets completed. The solid curves in first and second figures, for Daya-Bay, correspond to 1 and 2 MeV, respectively.  The solid curve for RENO and T2K correspond to 1.4  and 100 MeV, respectively. Dashed and dotted curves  correspond to lower and upper limits of neutrino energies for which the GP predictions for both signs of $\Delta_{31}$ are different. Blue and red curves correspond to predictions for positive and negative signs of $\Delta_{31}$, respectively.}
	\label{fig:endpoint}
\end{figure}

As the baseline length and energies for experiments such as  Daya-Bay, RENO and T2K allows at least one complete cycle of GP, measurement of GP of neutrinos with energy within specific energy range (such that condition for cluster point is satisfied) can enable resolving the sign ambiguity in $\Delta_{31}$. This can be seen from Fig.~\ref{fig:endpoint} where predictions for GP for both signs of  $\Delta_{31}$ are shown for various experimental set-ups. Here $L$ for these set-ups are fixed at their baseline length and $E$ is selected so that $L/E$ corresponds to one cycle of GP. It is obvious from the figure that the predictions for GP are different for + and - signs of $\Delta_{31}$ and hence one can determine the sign of $\Delta_{31} $. For Daya-Bay, the predictions for both signs of  $\Delta_{31}$ are different for all values of $CP$ violating phase for $E$ between   (0.95 - 1.02) MeV and   (1.97 - 2.04) MeV with $L$ fixed at the baseline length.

\FloatBarrier

 For RENO (T2K), the predictions for both signs of  $\Delta_{31}$ are different for neutrino energy range  (1.38 - 1.42) MeV   ((95 - 101) MeV). In Fig.~\ref{fig:endpoint} dashed and dotted curves represent the case of the lowest and the highest limits of these energy ranges, respectively. Solid curves correspond to that specific energy value at which a cycle of GP gets completed (eg. for RENO, solid curve correspond to E = 1.4 MeV). Hence measurement of GP would be helpful in resolving the degeneracy in $\Delta_{31}$. The points in Fig.~\ref{fig:endpoint} where all three curves representing different values of energy, for specific sign of $\Delta_{31}$, intersect because at the value of $CP$-phase, corresponding to intersection point (say $\delta \sim 195^o$ for RENO), the value of GP remains constant in the energy range mentioned above for different experimental set-ups.

\section{Geometric phase in terms of survival and oscillation probabilities}
	Experimental detection of GP usually makes use of interference set-ups. However, in the case of neutrinos, this would not be feasible \cite{rafflet,kim}. Nevertheless, one could envisage an interference experiment in energy space. This idea was used in \cite{pmehta} to show, in the context of two-flavor neutrino oscillations, that the topological (Pancharatnam) phase of the interference term was implicit in the transition probabilities associated with the neutrino oscillations. It was seen to be a consequence of the orthogonality of the two-flavor PMNS matrix. The analysis was made by exploiting the analogy between the geometry of two-state neutrino system and polarization states in optics. 
	
	The currently running experiments related to neutrino oscillations are designed to measure only survival ($P_{{\alpha} {\alpha}}$) or oscillation  ($P_{{\alpha}{\beta}}$) probabilities. Hence in order to measure geometric phase,  it should be expressed in terms of these experimentally measurable quantities. This can be  illustrated in a simple way by working within the context of two flavor neutrino oscillations. The probabilities are given by
	\begin{eqnarray}
	P_{ee} &=& 1 - \sin^2(2\theta) \, \sin^2(\frac{\Delta L}{4 E \hbar c}), \label{Psurv}
	\\
	P_{e\mu} &=& \sin^2(2\theta) \, \sin^2(\frac{\Delta L}{4 E \hbar c}).  \label{Posc}
	\end{eqnarray}
	Here, we have neglected the matter effect for simplicity.  In the context of two flavor neutrino oscillations, Eqs.(\ref{gp1}), (\ref{gp2}), (\ref{gp3}), (\ref{gp4})  reduce to
	\begin{eqnarray}
	\beta_{ee} &=& -\phi t \cos(2\theta) + \tan^{-1}[\cos(2\theta) \tan(\phi t)],\label{bee} \\
	\beta_{e\mu} &=& \phi t \cos(2\theta) -\frac{\pi}{2},\label{bemu} \\
	\beta_{\mu \mu} &=& \phi t \cos(2\theta) - \tan^{-1}[\cos(2\theta) \tan(\phi t)],\label{bmumu} \\
	\beta_{\mu e} &=& -\phi t \cos(2\theta) -\frac{\pi}{2},\label{bmue}
	\end{eqnarray}
	Using Eqs. (\ref{Psurv})- (\ref{bmue}),  the components of geometric phase can be written in terms of neutrino survival or oscillation probability and their average values: 	
		\begin{widetext}
		\begin{eqnarray}
		\beta_{ee}\big(P_{ee},\left<P_{ee}\right>\big) &=& -\frac{1}{2} \cos^{-1}\bigg[\frac{P_{ee}-\left<P_{ee}\right>}{1 - \left<P_{ee}\right>}\bigg]\sqrt{2\left<P_{ee}\right>-1}\nonumber \\
		&&+\tan^{-1}\bigg[\sqrt{2\left<P_{ee}\right>-1}~ \tan\bigg(\frac{1}{2} \cos^{-1}\bigg[\frac{P_{ee}-\left<P_{ee}\right>}{1-\left<P_{ee}\right>}\bigg]\bigg)\bigg],~~~~~
		\\  
		\beta_{e\mu}\big(P_{e\mu},\left<P_{e\mu}\right>\big) &=& \frac{1}{2} \cos^{-1}\bigg[1-\frac{P_{e\mu}}{\left<P_{e\mu}\right>}\bigg] \sqrt{1 - 2 \left<P_{e\mu}\right>} - \frac{\pi}{2},
		\\
		\beta_{\mu \mu}\big(P_{\mu \mu},\left<P_{\mu \mu}\right>\big) &=& \frac{1}{2} \cos^{-1}\bigg[\frac{P_{\mu \mu} - \left<P_{\mu \mu}\right>}{1 - \left<P_{\mu \mu}\right>}\bigg] \sqrt{2\left<P_{\mu \mu}\right> - 1}\nonumber \\
		&&- \tan^{-1}\bigg[\sqrt{2\left<P_{\mu \mu}\right> - 1}~ \tan\bigg(\frac{1}{2} \cos^{-1}\bigg[\frac{P_{\mu \mu} - \left<P_{\mu \mu}\right>}{1 - \left<P_{\mu \mu}\right>}\bigg]\bigg)\bigg],~~~~~
		\\
		\beta_{\mu e}\big(P_{\mu e},\left<P_{\mu e}\right>\big) &=& -\frac{1}{2} \cos^{-1}\bigg[1-\frac{P_{\mu e}}{\left<P_{\mu e}\right>}\bigg]\sqrt{1 - 2 \left<P_{\mu e}\right>} - \frac{\pi}{2}.
		\end{eqnarray}     
	    \end{widetext}    
	Here $\left<P_{\alpha \beta}\right>$ is the average value of probability $P_{\alpha \beta}$ and is given by
	\begin{equation}\nonumber
	 \left<P_{\alpha \beta}\right> =
	\begin{cases}
	1 - \frac{\sin^2(2\theta)}{2}  & {\rm for} ~~\alpha = \beta \\
	\frac{\sin^2(2\theta)}{2}     &   {\rm  for} ~~\alpha \neq \beta
	\end{cases}
	\end{equation}
Thus we see that the phases associated with neutrino oscillations can be expressed in terms of the experimentally measurable quantities.

\section{Conclusions}
We study non-cyclic GP, which would be, in principle, easier to observe than its cyclic counterpart, in the context of three flavor neutrino oscillations in the presence of matter and $CP$ violating effects at various man-made facilities, such as the reactor and accelerator neutrino experimental set-ups. The geometric phase is seen to be sensitive to the  sign ambiguity in $\Delta_{31}$. We analyze GP in the context of two  reactor neutrino experimental set-ups, Daya-Bay \& RENO and two accelerator experiments, T2K \& NO$\nu$A.  We find that for experimental facilities where the geometric phase can complete  atleast one cycle, all geometric phase curves corresponding  to different values of $CP$ violating phase,  converge to a single point, called the {\textit{cluster point}}. These cluster points are distinct for positive and negative signs of $\Delta_{31}$. Thus experimental set-ups, such as  T2K, Daya-Bay and RENO where atleast one complete cycle of GP is possible, could help in resolving the neutrino mass hierarchy problem. The normal and inverted types of mass hierarchy are distinguishable for CP-violating phase $\delta \in [0,2\pi]$ in the suggested energy range for all of these three experiments.


\begin{thebibliography}{25}
	
	\bibitem{Bahcall:2004ut} 
	J.~N.~Bahcall, M.~C.~Gonzalez-Garcia and C.~Pena-Garay,
	JHEP {\bf 0408},  016 (2004).
	
	\bibitem{Eguchi:2002dm}
	K.~Eguchi {\it et al.} [KamLAND Collaboration],
	Phys.\ Rev.\ Lett. {\bf 90}, 021802 (2003).
	
	\bibitem{Araki:2004mb} 
	T.~Araki {\it et al.}  [KamLAND Collaboration],
	Phys.\ Rev.\ Lett.\  {\bf 94},   081801 (2005).
	
	\bibitem{Ashie:2004mr} 
	Y.~Ashie {\it et al.}  [Super-Kamiokande Collaboration],
	Phys.\ Rev.\ Lett.\  {\bf 93},  101801 (2004).
	
	\bibitem{Michael:2006rx} 
	D.~G.~Michael {\it et al.}  [MINOS Collaboration],
	Phys.\ Rev.\ Lett.\  {\bf 97},   191801 (2006).
	
	
	\bibitem{Abe:2013hdq} 
	K.~Abe {\it et al.}  [T2K Collaboration],
	Phys.\ Rev.\ Lett.\  {\bf 112}, 061802 (2014).
	
	\bibitem{Abe:2013fuq} 
	K.~Abe {\it et al.}  [T2K Collaboration],
	Phys.\ Rev.\ Lett.\  {\bf 111}, no. 21  211803 (2013).
	
	\bibitem{Blasone:2009xk} 
  M.~Blasone, A.~Capolupo, E.~Celeghini and G.~Vitiello,
  Phys.\ Lett.\ B {\bf 674}, 73 (2009)
  [arXiv:0903.1578 [hep-th]].
	
	
	\bibitem{Blasone:2014jea} 
	M.~Blasone, F.~Dell'Anno, S.~De Siena and F.~Illuminati,
	Europhys.\ Lett.\  {\bf 106}, 30002 (2014)
	[arXiv:1401.7793 [quant-ph]].
	
	\bibitem{Alok:2014gya} 
	A.~K.~Alok, S.~Banerjee and S.~U.~Sankar,
	Nucl.\ Phys.\ B {\bf 909}, 65 (2016)
	[arXiv:1411.5536 [hep-ph]].
	
	\bibitem{Banerjee:2015mha} 
	S.~Banerjee, A.~K.~Alok, R.~Srikanth and B.~C.~Hiesmayr,
	Eur.\ Phys.\ J.\ C {\bf 75}, no. 10, 487 (2015)
	[arXiv:1508.03480 [hep-ph]].


	\bibitem{Naikoo:2017fos} 
	J.~Naikoo, A.~K.~Alok, S.~Banerjee, S.~U.~Sankar, G.~Guarnieri and B.~C.~Hiesmayr,
	arXiv:1710.05562 [hep-ph].
	
	\bibitem{pdg}
	K.A. Olive et al. (Particle Data Group), Chin. Phys. C {\bf 38}, 090001 (2014).
	
	
	
	\bibitem{Dick}
	K.~Dick, M.~Freund, M.~Lindner and A.~Romanino,
	Nucl.\ Phys.\ B {\bf 562}, 29 (1999)
	[hep-ph/9903308].
	
	
	\bibitem{Blasone}
	M.~Blasone and G.~Vitiello,
	Phys.\ Rev.\ D {\bf 60}, 111302 (1999)
	[hep-ph/9907382].
	
	\bibitem{He}
	X.~G.~He, X.~Q.~Li, B.~H.~J.~McKellar and Y.~Zhang,
	Phys.\ Rev.\ D {\bf 72}, 053012 (2005)
	[hep-ph/0412374].
	
	\bibitem{wilczek}
	A. Shapere and F. Wilczek, {\it Geometric Phases in Physics} (World Scientific, Singapore, 1989).
	
	\bibitem{Xiang}
	X.~B.~Wang, L.~C.~Kwek, Y.~Liu and C.~H.~Oh,
	Phys.\ Rev.\ D {\bf 63}, 053003 (2001)
	[hep-ph/0006204].
	
	
	\bibitem{Dajka}
	J.~Dajka, J.~Syska and J.~Luczka,
	Phys.\ Rev.\ D {\bf 83}, 097302 (2011)
	[arXiv:1309.7628 [hep-ph]].
	
	\bibitem{Capolupo}
	A.~Capolupo, S.~M.~Giampaolo, B.~C.~Hiesmayr and G.~Vitiello,
	Phys.\ Lett.\ B {\bf 780}, 216 (2018)
	arXiv:1610.08679 [hep-ph].
	
	\bibitem{pmehta}
	P.~Mehta, Phys.\ Rev.\ D\  {\bf 79}, 096013 (2009).	
	
	\bibitem{Johns:2016wjd} 
	L.~Johns and G.~M.~Fuller,
	Phys.\ Rev.\ D {\bf 95}, no. 4, 043003 (2017)
	[arXiv:1612.06940 [hep-ph]].
	
	
	\bibitem{berry} M. V. Berry, Proc. R. Soc. London, Ser. A {\bf 392}, 45 (1984).
	
	\bibitem{pancham} S. Pancharatnam, Proc. Indian Acad. Sci. A {\bf 44}, 247
	(1956).
	
	\bibitem{nityananda} S. Ramaseshan and R. Nityananda, Curr. Sci. {\bf 55}, 1225
	(1986).
	
	\bibitem{simon} B. Simon, Phys. Rev. Lett. {\bf 51}, 2167 (1983).
	
	\bibitem{aharanov} Y. Aharonov and J. Anandan, Phys. Rev. Lett. {\bf 58}, 1593 (1987).
	
	\bibitem{samuel} J. Samuel and R. Bhandari, Phys. Rev. Lett. {\bf 60}, 2339 (1988).
	
	\bibitem{tong} D. M. Tong, E. Sj$\ddot{o}$qvist, L. C. Kwek and C. H. Oh, Phys.
	Rev. Lett. {\bf 93}, 080405 (2004).
	
	\bibitem{zenardi} A. T. Rezakhani and P. Zanardi, Phys. Rev. A {\bf 73}, 052117 (2006).
	
	\bibitem{villar} F. C. Lombardo, P. I. Villar, Phys. Rev. A {\bf 74}, 042311 (2006).
	
	\bibitem{sb1} S. Banerjee and R. Srikanth, Eur. Phys. J. D {\bf 46}, 335 (2008).
	
	\bibitem{sb2} S. N. Sandhya and S. Banerjee, Euro. Phys. J. D {\bf 66}, 168 (2012).
	
	\bibitem{sb3} S. Banerjee, C. M. Chandrashekar and A. K. Pati, Phys. Rev. A {\bf 87}, 042119 (2013).
	
	\bibitem{lidar} M. S. Sarandy and D. A. Lidar, Phys. Rev. A {\bf 73}, 062101 (2006).
	
	\bibitem{Capolupo1} 
	A. Capolupo
	Phys.\ Rev.\ D {\bf 84}, 116002 (2011)
	arXiv:1112.1337 [hep-ph].
	
	
	
	\bibitem{T01} 
  T.~Ohlsson and H.~Snellman,
  J.\ Math.\ Phys.\  {\bf 41}, 2768 (2000)
  Erratum: [J.\ Math.\ Phys.\  {\bf 42}, 2345 (2001)]
  [hep-ph/9910546].
	
	
	\bibitem{Tommy}
	T.~Ohlsson and H.~Snellman,
	Phys.\ Lett.\ B {\bf 474}, 153 (2000)
	Erratum: [Phys.\ Lett.\ B {\bf 480}, 419 (2000)]
	[hep-ph/9912295].
	
	\bibitem{T02}
  T.~Ohlsson,
  Phys.\ Scripta T {\bf 93}, 18 (2001).
    
	
		
	\bibitem{An}
	F.~P.~An {\it et al.} [Daya Bay Collaboration],
	Phys.\ Rev.\ Lett.\  {\bf 112}, 061801 (2014)
	[arXiv:1310.6732 [hep-ex]].
	
	\bibitem{Balantekin}
	F.~P.~An {\it et al.} [Daya Bay Collaboration],
	Phys.\ Rev.\ Lett.\  {\bf 115}, no. 11, 111802 (2015)
	[arXiv:1505.03456 [hep-ex]].
	
	\bibitem{Patterson}
	R.~B.~Patterson [NOvA Collaboration],
	Nucl.\ Phys.\ Proc.\ Suppl.\  {\bf 235-236}, 151 (2013)
	[arXiv:1209.0716 [hep-ex]].
	
	
	\bibitem{Adamson}
	P.~Adamson {\it et al.} [NOvA Collaboration],
	Phys.\ Rev.\ Lett.\  {\bf 116}, no. 15, 151806 (2016)
	[arXiv:1601.05022 [hep-ex]].
	
	
	\bibitem{rafflet}
	G. G. Raffelt, {\it Stars as Laboratories for Fundamental Physics: The Astrophysics of Neutrinos, Axions, and Other Weakly Interacting Particles} (University of Chicago Press, Chicago, 1996).
	
	\bibitem{kim}
	C. Giunti and C. W. Kim, {\it Fundamentals of Neutrino Physics and Astrophysics} (Oxford University Press, 2007).
	
\end{thebibliography}
\end{document}